\documentclass[onecolumn,draft]{IEEEtran}
\usepackage{epsfig}
\usepackage{amsmath}
\usepackage{amstext}
\usepackage{amsfonts}
\usepackage{amssymb}
\usepackage{eucal}
\usepackage{graphicx}
\usepackage{verbatim}
\usepackage{bm}
\usepackage{flushend}

\usepackage{tikz}
\usepackage{pgfplots}

\usetikzlibrary{arrows,petri,topaths}
\usepackage{tkz-berge}

\newcommand{\RR}{{\mathbb{R}}}
\newcommand{\NN}{{\mathbb{N}}}

\newcommand{\CC}{{\mathbb{C}}}

\newcommand{\trans}{{\sf T}}

\newcommand{\blambda}{{\bm \lambda}}

\newcommand{\oh}{{\frac{1}{2}}}

\newcommand{\asto}{\overset{\rm a.s.}{\longrightarrow}}

\newcommand{\EE}{{\rm E}}

\DeclareMathOperator{\tr}{tr}
\DeclareMathOperator{\diag}{diag}

\newcounter{ctheorem}
\newtheorem{theorem}[ctheorem]{Theorem}

\newcounter{cproposition}
\newtheorem{proposition}[cproposition]{Proposition}
\newcounter{ccorollary}
\newtheorem{corollary}[ccorollary]{Corollary}
\newcounter{clemma}
\newtheorem{lemma}[clemma]{Lemma}
\newcounter{cdefinition}
\newtheorem{definition}[cdefinition]{Definition}
\newcounter{cremark}
\newtheorem{remark}[cremark]{Remark}

\begin{document}
\bibliographystyle{IEEEtran}

\title{Robust Estimates of Covariance Matrices in the Large Dimensional Regime}

\author{Romain~Couillet$^{1}$, Fr\'ed\'eric Pascal$^2$, and Jack W. Silverstein$^3$\thanks{Silverstein's work is supported by the U.S. Army Research Office, Grant W911NF-09-1-0266. Couillet's work is supported by the ERC MORE EC--120133.} \\ {\it $^1$ Telecommunication department, Sup\'elec, Gif sur Yvette, France.} \\ {\it $^2$ SONDRA Laboratory, Sup\'elec, Gif sur Yvette, France.} \\ {\it $^3$ Department of Mathematics, North Carolina State University, NC, USA.}}
\maketitle

\begin{abstract}
This article studies the limiting behavior of a class of robust population covariance matrix estimators, originally due to Maronna in 1976, in the regime where both the number of available samples and the population size grow large. Using tools from random matrix theory, we prove that, for sample vectors made of independent entries having some moment conditions, the difference between the sample covariance matrix and (a scaled version of) such robust estimator tends to zero in spectral norm, almost surely. This result can be applied to various statistical methods arising from random matrix theory that can be made robust without altering their first order behavior.
\end{abstract}

\section{Introduction}
\label{sec:intro}

Many multi-variate signal processing detection and estimation techniques are based on the empirical covariance matrix of a sequence of samples $x_1,\ldots,x_n$ from a random population vector $x\in\CC^N$. Assuming $\EE[x]=0$ and $\EE[xx^*]=C_N$, the strong law of large numbers ensures that, for independent and identically distributed (i.i.d.) samples,
\begin{align*}
\hat{S}_N=\frac1n\sum_{i=1}^nx_ix_i^*\to C_N
\end{align*}
almost surely (a.s.), as the number $n$ of samples increases. Many subspace methods, such as the multiple signal classifier (MUSIC) algorithm and its derivatives \cite{SCH86,SCH91}, heavily rely on this property by identifying $C_N$ with $\hat{S}_N$, leading to appropriate approximations of functionals of $C_N$ in the large $n$ regime. However, this standard approach has two major limitations: the inherent inadequacy to small sample sizes (when $n$ is not too large compared to $N$) and the lack of robustness to outliers or heavy-tailed distribution of $x$. Although the former issue was probably the first historically recognized, it is only recently that significant advances have been made using random matrix theory \cite{MES08}. As for the latter, it has spurred a strong wave of interest in the seventies, starting with the works from Huber \cite{HUB64} on robust M-estimation. The objective of this article is to provide a first bridge between the two disciplines by introducing new fundamental results on robust M-estimates in the random matrix regime where both $N$ and $n$ grow large at the same rate.

Aside from its obvious simplicity of analysis, the {\it sample covariance matrix} (SCM) $\hat{S}_N$ is an object of primal interest since it is the maximum likelihood estimator of $C_N$ for $x$ Gaussian. When $x$ is not Gaussian, the SCM as an approximation of $C_N$ may however perform very poorly. This problem was identified in multiple areas such as multivariate signal processing or financial asset management, but was particularly recognized in adaptive radar and sonar processing where the signals under study are characterized by impulsive noise and outlying data. Robust estimation theory aims at tackling this problem \cite{MAR06}. Among other solutions, the so-called robust M-estimators of the population covariance matrix, originally introduced by Huber \cite{HUB64} and investigated in the seminal work of Maronna \cite{MAR76}, have imposed themselves as an appealing alternative to the SCM. This estimator, which we denote $\hat{C}_N$, is defined implicitly as a solution of\footnote{Our expression differs from the standard convention where $x_i^*\hat{C}_N^{-1}x_i$ is traditionally not scaled by $1/N$. The current form is however more convenient for analysis in the large $N,n$ regime.}
\begin{align}
	\label{def:hatCN}
	\hat{C}_N = \frac1n\sum_{i=1}^n u\left( \frac1Nx_i^*\hat{C}_N^{-1}x_i\right) x_ix_i^*
\end{align}
for $u$ a nonnegative function with specific properties. These estimators are particularly appropriate as they are the maximum likelihood estimates of $C_N$ for specific distributions of $x$ and some specific choices of $u$, such as the family of elliptical distributions \cite{KEL70}. For any such $u$, $\hat{C}_N$ is, up to a scalar, a consistent estimate for $C_N$ for $N$ fixed and $n\to \infty$, see e.g. \cite{OLI12}. The robust estimators are also used to cope with distributions of $x$ with heavy tails or showing a tendency to produce outliers, such as when $\Vert x\Vert^2$ has a K-distribution often met in the context of adaptive radar processing with impulsive clutter \cite{WAT85}. In this article, the concept of robustness is to be understood along this general theory.

A second angle of improvement of subspace methods has recently emerged due to advances in random matrix theory. The latter aims at studying the statistical properties of matrices in the regime where both $N$ and $n$ grow large. It is known in particular that, if $x=A_N y$ with $y\in\CC^M$, $M\geq N$, a vector of independent entries with zero mean and unit variance, then, under some conditions on $C_N=A_NA_N^*$ and $y$, in the large $N,n$ (and $M$) regime, the eigenvalue distribution of (almost every) $\hat{S}_N$ converges weakly to a limiting distribution described implicitly by its Stieltjes transform \cite{SIL95b}. When $C_N$ is the identity matrix for all $N$, this distribution takes an explicit form known as the Mar\u{c}enko-Pastur law \cite{MAR67}. Under some additional moment conditions on the entries of $y$, it has also been shown that the eigenvalues of $\hat{S}_N$ cannot lie infinitely often away from the support of the limiting distribution \cite{SIL98}. In the past ten years, these two results and subsequent works have been applied to revisit classical signal processing techniques such as signal detection schemes \cite{BIA10} or subspace methods \cite{MES08b,COU10b}. In these works, traditional {\it $n$-consistent} detection and estimation methods were improved into {\it $(N,n)$-consistent} approaches, i.e. they provide estimates that are consistent in the large $N,n$ regime rather than in the fixed $N$ and large $n$ regime. These improved estimators are often referred to as G-estimators.

In this article, we study the asymptotic first order properties of the robust M-estimate $\hat{C}_N$ of $C_N$, given by \eqref{def:hatCN}, in the regime where $N$, $n$ (and $M$) grow large simultaneously, hereafter referred to as the random matrix regime. Although the study of the SCM $\hat{S}_N$ for vectors $x$ with rather general distributions is accessible to random matrix theory, as in e.g. the case of elliptical distributions \cite{ELK09}, the equivalent analysis for $\hat{C}_N$ is often very challenging. In the present article, we restrict ourselves to vectors $x$ of the type $x=A_Ny$ with $y$ having independent zero-mean entries. One important technical challenge brought by the matrix $\hat{C}_N$, usually not met in random matrix theory, lies in the dependence structure between the vectors $\{u(\frac1Nx_i^*\hat{C}_N^{-1}x_i)^\frac12x_i\}_{i=1}^n$ (as opposed to the independent vectors $\{x_i\}_{i=1}^n$ for the matrix $\hat{S}_N$). We fundamentally rely on the set of assumptions on the function $u$ taken by Maronna in \cite{MAR76} to overcome this difficulty. Our main contribution consists in showing that, in the large $N,n$ regime, and under some mild assumptions, $\Vert \hat{C}_N-\alpha\hat{S}_N\Vert \to 0$, a.s., for some constant $\alpha>0$ dependent only on $u$. This result is in particular in line with the conjecture made in \cite{FRA08} according to which $\Vert \hat{C}_N-\alpha\hat{S}_N\Vert \asto 0$ for the function $u(s)=1/s$ studied extensively by Tyler \cite{TYL88,KEN91}; however, the function $u(s)=1/s$ does not enter our present scheme as it creates additional difficulties which leave the conjecture open.  

A major practical consequence of our result is that the matrix $\hat{S}_N$, at the core of many random matrix-based estimators, can be straightforwardly replaced by $\hat{C}_N$ without altering the first order properties of these estimators. We generically call the induced estimators {\it robust G-estimators}. As an application example, we shall briefly introduce an application to robust direction-of-arrival estimation accounting for large $N,n$ based on the earlier estimator \cite{MES08c}.

The remainder of the article is structured as follows. Section~\ref{sec:results} provides our theoretical results along with an application to direction-of-arrival estimation. Section~\ref{sec:conclusion} then concludes the article. All technical proofs are detailed in the appendices.

{\it Notations:} The arrow `$\asto$' denotes almost sure convergence. For $A\in\CC^{N\times N}$ Hermitian, $\lambda_1(A)\leq \ldots \leq \lambda_N(A)$ are its ordered eigenvalues. The norm $\Vert \cdot \Vert$ is the spectral norm for matrices and the Euclidean norm for vectors. For $A,B$ Hermitian, $A\succeq B$ means that $A-B$ is nonnegative definite. The notation $A^*$ denotes the Hermitian transpose of $A$. We also write $\imath=\sqrt{-1}$.

\section{Main results}
\label{sec:results}

\subsection{Theoretical results}

Let $X=[x_1,\ldots,x_n]\in\CC^{N\times n}$, where $x_i=A_Ny_i\in\CC^N$, with $y_i=[y_{i1},\ldots,y_{iM}]^\trans\in\CC^M$ having independent entries with zero mean and unit variance, $A_N\in\CC^{N\times M}$, and $C_N\triangleq A_NA_N^*\in\CC^{N\times N}$ be a positive definite matrix. We denote $c_N\triangleq N/n$, $\bar{c}_N\triangleq M/N$, and define the sample covariance matrix $\hat{S}_N$ of the sequence $x_1,\ldots,x_n$ by
\begin{equation*}
	\hat{S}_N\triangleq \frac1nXX^*=\frac1n\sum_{i=1}^nx_ix_i^*.
\end{equation*}


Let $u:\RR^+\to\RR^+$ ($\RR^+=[0,\infty)$) be a function fulfilling the following conditions:
\begin{itemize}
	\item[(i)] $u$ is nonnegative, nonincreasing, and continuous on $\RR^+$;
	\item[(ii)] the function $\phi: \RR^+\to \RR^+,~s\mapsto su(s)$  is nondecreasing and bounded, with $\sup_{x}\phi(x) = \phi_\infty>1$. Moreover, $\phi$ is increasing in the interval where $\phi(s)<\phi_\infty$.
\end{itemize}

Classical M-estimators $\hat{C}_N$ defined by \eqref{def:hatCN} for such function $u$ include the Huber estimator, with $\phi(s)=\frac{\phi_\infty}{\phi_\infty-1}s$ for $s\in[0,\phi_\infty-1]$, $\phi_\infty>1$, and $\phi(s)=\phi_\infty$ for $s\geq \phi_\infty-1$. Since $u(s)$ is constant for $s\leq \phi_\infty-1$ and decreases for $s\geq \phi_\infty-1$, this estimator weights the majority of the samples $x_1,\ldots,x_n$ by a common factor and reduces the impact of the outliers. The widely used function $u(s)=(1+t)(t+x)^{-1}$ for some $t>0$ shows similar properties, here with $\phi_\infty=1+t$.\footnote{Note that this function intervenes in the maximum-likelihood estimator of the scatter matrix of Student-t distributed random vectors \cite{OLI12}. Here we do not make any such maximum-likelihood consideration for the selection of $u$.} Other classical $u$ functions, adapted to specific distributions of the samples, can be found in the survey \cite{OLI12}. In any of these scenarios, robustness can be controlled by properly setting $\phi_\infty$.

To pursue, we need the following statistical assumptions on the large dimensional random matrices under study.

\vspace{0.2cm}
{\bf A1.} The random variables $y_{ij}$, $i\leq n$, $j\leq M$, are independent either real or circularly symmetric complex (i.e. $\EE[y_{ij}^2]=0$) with $\EE[y_{ij}]=0$ and $\EE[|y_{ij}|^2]=1$. Also, there exists $\eta>0$ and $\alpha>0$, such that, for all $i,j$, $\EE[|y_{ij}|^{8+\eta}]<\alpha$. 

%

\vspace{0.2cm}
{\bf A2.} $\bar{c}_N\geq 1$ and, as $n\to\infty$,
\begin{align*}
	0<\lim\inf_n c_N\leq &\lim\sup_n c_N<1, \quad \lim\sup_n \bar{c}_n<\infty.
\end{align*}


\vspace{0.2cm}
{\bf A3.} There exists $C_-,C_+>0$ such that 
\begin{align*}
	C_-<\lim\inf_N \{\lambda_1(C_N)\}\leq \lim\sup_N \{\lambda_N(C_N)\}< C_+.
\end{align*}

Note that the assumptions neither request the entries of $y$ to be identically distributed nor impose the existence of a continuous density. This assumption is adequate for a large range of application scenarios such as factor models in finance or general signal processing models with independent entry-wise non-Gaussian noise (e.g. distributed antenna array processing), although the requirement of independence in the entries of $y$ is somewhat uncommon in the classical applications of robust estimation theory. The entry-wise independence is however central in this article for the emergence of a concentration of the quadratic forms $\frac1Nx_i^*\hat{C}_N^{-1}x_i$, $i=1,\ldots,n$. Further generalizations, e.g. to elliptical distributions for $x$, would break this effect and would certainly entail a much different asymptotic behavior of $\hat{C}_N$. These important considerations are left to future work. 

Technically, {\bf A1}--{\bf A3} mainly ensure that the eigenvalues of $\hat{S}_N$ and $\hat{C}_N$ lie within a compact set away from zero, a.s., for all $N,n$ large, which is a consequence (although non immediate) of \cite{SIL98,COU10b}. Note also that {\bf A2} demands $\lim\inf_N c_N>0$, so that the following results {\it do not} contain the results from \cite{MAR76,KEN91}, in which $N$ is fixed and $n\to\infty$, as special cases. With these assumptions, we are now in position to provide the main technical result of this article. 

\vspace{0.2cm}
\begin{theorem}
	\label{th:1}
	Assume {\bf A1}--{\bf A3} and consider the following matrix-valued fixed-point equation in $Z\in\CC^{N\times N}$,
	\begin{align}
		\label{eq:hatCN}
		Z = \frac1n\sum_{i=1}^n u\left(\frac1N x_i^*Z^{-1}x_i \right)x_ix_i^*.
	\end{align}
	Then, we have the following results. 
	\begin{itemize}
		\item[(I)] There exists a unique solution to \eqref{eq:hatCN} for all large $N$ a.s. We denote $\hat{C}_N$ this solution, defined as
			\begin{align*}
				\hat{C}_N = \lim_{t\to\infty} Z^{(t)}
			\end{align*}
			where $Z^{(0)}=I_N$ and, for $t\in\NN$,
	\begin{align*}
		Z^{(t+1)} = \frac1n\sum_{i=1}^n u\left(\frac1N x_i^*(Z^{(t)})^{-1}x_i \right)x_ix_i^*.
	\end{align*}
\item[(II)] Defining $\hat{C}_N$ arbitrarily when \eqref{eq:hatCN} does not have a unique solution, we also have
	\begin{align*}
		\left\Vert \phi^{-1}(1)\hat{C}_N - \hat{S}_N \right\Vert \asto 0.
	\end{align*}
	\end{itemize}
\end{theorem}
\begin{IEEEproof}
	The proof is provided in Appendix \ref{app:th1}.
\end{IEEEproof}
\vspace{0.2cm}

An immediate corollary of Theorem~\ref{th:1} is the asymptotic closeness of the ordered eigenvalues of ${\phi^{-1}(1)}\hat{C}_N$ and $\hat{S}_N$.

\vspace{0.2cm}
\begin{corollary}
	\label{co:spacing}
	Under the assumptions of Theorem~\ref{th:1},
	\begin{align*}
		\max_{i\leq N} \left| {\phi^{-1}(1)}\lambda_i(\hat{C}_N) - \lambda_i(\hat{S}_N) \right| &\asto 0. 
	\end{align*}
\end{corollary}
\begin{IEEEproof}
	The proof is provided in Appendix~\ref{app:th1}.
\end{IEEEproof}
\vspace{0.2cm}

Some comments are called for to understand Theorem~\ref{th:1} in the context of robust M-estimation.

Theorem~\ref{th:1}--(I) can be first compared to the result from Maronna \cite[Theorem~1]{MAR76} which states that a solution to \eqref{eq:hatCN} exists for each set $\{x_1,\ldots,x_n\}$ under certain conditions on the dimension of the space spanned by the $n$ vectors, as well as on $u(s)$, $N$, and $n$ (in particular $u(s)$ must satisfy $\phi_\infty>n/(n-N)$ in \cite{MAR76}). Our result may be considered more interesting in practice in the sense that the system sizes $N$ and $n$ no longer condition $\phi_\infty$ and therefore do not constrain the definition of $u(s)$. Theorem~\ref{th:1}--(I) can also be compared to the results on uniqueness \cite{MAR76,KEN91} which hold for all $N,n$ under some further conditions on $u(s)$, such as $\phi(s)$ is strictly increasing \cite{MAR76}. The latter assumption is particularly demanding as it may reject some M-estimators such as the Huber M-estimator for which $\phi(s)$ is constant for large $s$. Theorem~\ref{th:1}--(I) trades these assumptions against a requirement for $N$ and $n$ to be ``sufficiently large'' and for $\{x_1,\ldots,x_n\}$ to belong to a probability one sequence. Precisely, we demand that there exists an integer $n_0$ depending on the random sequence $\{(x_1,\ldots,x_n)\}_{n=1}^\infty$, such that for all $n\geq n_0$, existence and uniqueness are established under no further condition than the definition (i)--(ii) of $u(s)$ and {\bf A1}--{\bf A3}. 

Theorem~\ref{th:1}--(II), which is our main result, states that, as $N$ and $n$ grow large with a non trivial limiting ratio, the fixed-point solution $\hat{C}_N$ (either always defined under the assumptions of \cite{MAR76,KEN91} or defined a.s. for large enough $N$) is getting asymptotically close to the sample covariance matrix, up to a scaling factor. This implies in particular that, while $\hat{C}_N$ is an $n$-consistent estimator of (a scaled version of) $C_N$ for $n\to\infty$ and $N$ fixed, in the large $N,n$ regime it has many of the same first order statistics as $\hat{S}_N$. This suggests that many results holding for $\hat{S}_N$ in the large $N,n$ regime should also hold for $\hat{C}_N$, at least concerning first order convergence. For instance, as will be seen through Corollary~\ref{co:RG-MUSIC}, one expects consistent estimators (in the large $N,n$ regime) based on functionals of $\hat{S}_N$ to remain consistent when using $\phi^{-1}(1)\hat{C}_N$ in place of $\hat{S}_N$ in the expression of the estimator. However, it is important to note that, in general, one cannot say much on second order statistics, i.e. regarding the comparison of the asymptotic performance of both estimators. The matrices $\hat{C}_N$, parametrizable through $u$, should then be seen as a class of alternatives for $\hat{S}_N$ which may possibly improve estimators based on $\hat{S}_N$ in the large (but finite) $N,n$ regime. Note also that Theorem~\ref{th:1} is independent of the choice of the distribution of the entries of $y$ (as long as the moment conditions are satisfied) or of the choice of the function $u$, which is in this sense similar to the equivalent result in the classical fixed-$N$ large-$n$ regime \cite{OLI12}.

	In a similar context, it is shown in \cite{SIL98} and \cite{YIN88b} that the eigenvalues of $\hat{S}_N$ are asymptotically contained in the support of their limiting compactly supported distribution if and only if the entries of $y$ have finite fourth order moment. This first suggests that the technical assumption {\bf A1} which requires $y$ to have uniformly bounded $8+\eta$ moment may be relaxed to $y_{ij}$ having only finite fourth order moments for Theorem~\ref{th:1} to hold. This being said, since most of the aforementioned $(N,n)$-consistent estimators involving $\hat{C}_N$ or $\hat{S}_N$ rely on a non-degenerate behavior of these eigenvalues (see e.g. \cite[Chapters~16--17]{COUbook} for details), the finite fourth order moment condition cannot possibly be further relaxed for these estimators to be usable. As a consequence, although {\bf A1} might seem very restrictive in a robust estimation framework as it discards the possibility to consider distributions of $x$ with heavy tail behavior, it is a close to necessary condition for robust estimation in the random matrix regime to be meaningful.

In terms of applications to signal processing, recall first that the $n$-consistency results on robust estimation \cite{MAR76,KEN91} imply that many metrics based on functionals of $C_N$ can be consistently estimated by replacing $C_N$ by $\hat{C}_N$. The inconsistency of the sample covariance matrix to the population covariance in the random matrix regime, along with Theorem~\ref{th:1}, suggest instead that this approach will lead in general to inconsistent estimators in the large $N,n$ regime, and therefore to inaccurate estimates for moderate values of $N,n,M$. However, any metric based on $C_N$, and for which an $(N,n)$-consistent estimator involving $\hat{S}_N$ exists, is very likely to be $(N,n)$-consistently estimated by replacing $\hat{S}_N$ by $\phi^{-1}(1)\hat{C}_N$. The interest of this replacement obviously lies in the possibility to improve the metric through an appropriate choice of $u$, in particular when $y$ exhibits outlier behavior or has heavy tails. 

\subsection{Application example}

A specific example can be found in the context of MUSIC-like estimation methods for array processing. In this example, $K$ signal sources imping on a collection of $N$ collocated sensors with angles of arrival $\theta_1,\ldots,\theta_K$. The data $x_i\in\CC^N$ received at time $i$ at the array is modeled as
\begin{align*}
	x_i = \sum_{k=1}^K \sqrt{p_k} s(\theta_k) z_{k,i} + \sigma w_i
\end{align*}
where $s(\theta)\in\CC^N$ is the deterministic unit norm steering vector for signals impinging the sensors at angle $\theta$, $z_{k,t}\in\CC$ is the signal source modeled as a zero mean, unit variance, and finite $8+\eta$ order moment random variable, i.i.d. across $t$ and independent across $k$, $p_k>0$ is the transmit power of source $k$ ($p_k<p_{\rm max}$ for some $p_{\rm max}>0$) and $\sigma w_i\in\CC^N$ is the received noise at time $t$, independent across $t$, with i.i.d. zero mean, variance $\sigma^2>0$, and finite $8+\eta$ order moment entries. Write $x_i=A_N y_i$, with $A_N \triangleq [S(\Theta)P^\frac12,\sigma I_N]$, $S(\Theta)=[s(\theta_1),\ldots,s(\theta_K)]$, $P=\diag(p_1,\ldots,p_K)$, and $y_i=(z_{1,t},\ldots,z_{K,t},w_i^\trans)^\trans\in\CC^{N+K}$. Then, with $N,n$ large and $K$ finite, Assumptions~{\bf A1}--{\bf A3} are met and Theorem~\ref{th:1} can be applied. This yields the following corollary of Theorem~\ref{th:1}.
\vspace{0.2cm}
\begin{corollary}[Robust G-MUSIC]
	\label{co:RG-MUSIC}
	Denote $E_W\in\CC^{N\times (N-K)}$ a matrix containing in columns the eigenvectors of $C_N$ with eigenvalue $\sigma^2$ and $\hat{e}_k$ the eigenvector of $\hat{C}_N$ with eigenvalue $\hat\lambda_k\triangleq \lambda_k(\hat{C}_N)$ (recall that $\hat\lambda_1\leq\ldots\leq \hat\lambda_N$), with $\hat{C}_N$ defined as in Theorem~\ref{th:1}. Then, as $N,n\to \infty$ in the regime of Assumption~{\bf A2}, and $K$ fixed, 
	\begin{align*}
		\gamma(\theta) - \hat{\gamma}(\theta) \asto 0
	\end{align*}
	where
	\begin{align*}
		\gamma(\theta) &= s(\theta)^* E_WE_W^* s(\theta) \\
		\hat{\gamma}(\theta) &= \sum_{i=1}^N \beta_i s(\theta)^* \hat{e}_i \hat{e}_i^* s(\theta)
\end{align*}
and
\begin{align*}
\beta_i &= \left\{
\begin{array}{ll}
	1+\sum_{k=N-K+1}^N \left( \frac{\hat\lambda_k}{\hat\lambda_i - \hat\lambda_k} - \frac{\hat\mu_k}{\hat\lambda_i-\hat\mu_k} \right) &,~i\leq N-K \\
	- \sum_{k=1}^{N-K} \left( \frac{\hat\lambda_k}{\hat\lambda_i - \hat\lambda_k} - \frac{\hat\mu_k}{\hat\lambda_i-\hat\mu_k} \right) &,~i>N-K
\end{array}
\right.
\end{align*}
with $\hat\mu_1\leq\ldots\leq \hat\mu_N$ the eigenvalues of $\diag(\hat\blambda)-\frac1n \sqrt{\hat\blambda}\sqrt{\hat\blambda}^\trans$, $\hat\blambda=(\hat\lambda_1,\ldots,\hat\lambda_N)^\trans$.
\end{corollary}
\begin{IEEEproof}
	The Corollary is exactly the algorithm \cite{MES08b} with $\hat{S}_N$ replaced by $\hat{C}_N$. The validity of this operation is proved in Appendix \ref{app:RG-MUSIC}.
\end{IEEEproof}
\vspace{0.2cm}

The function $\gamma(\theta)$ is the defining metric for the MUSIC algorithm \cite{SCH86}, the zeros of which contain the $\theta_i$, $i\in\{1,\ldots,K\}$. Corollary~\ref{co:RG-MUSIC} proves that the $N,n$-consistent G-MUSIC estimator of $\gamma(\theta)$ proposed by Mestre in \cite{MES08b} can be extended into a robust G-MUSIC method. The latter merely consists in replacing the sample covariance matrix $\hat{S}_N$ as in \cite{MES08b} by the robust estimator $\hat{C}_N$. The angles $\theta_i$ are then estimated as the deepest minima of $\hat{\gamma}(\theta)$. This technique can be seen through simulations to perform better than either MUSIC or G-MUSIC in the finite $(N,n)$ regime in the case of impulsive noise in the sense of {\bf A1}, for an appropriate choice of the function $u$. However, proving so requires the study of the second order statistics of $\gamma(\theta)$, which goes beyond the reach of the present article and is left to future work.

\section{Conclusion}
\label{sec:conclusion}

We have proved that a large family of robust estimates of population covariance matrices is consistent with the sample covariance matrix in the regime of both large population $N$ and sample $n$ sizes, this being valid irrespective of the sample distribution. This result opens up a new area of research for robust estimators in the random matrix regime. The results can be applied to improve a variety of signal processing techniques relying on random matrix methods but not accounting for noise impulsiveness yet. The exact performance gain of such improved methods however often relies on second order statistics which will be investigated in future work.

\appendices

\section{Proof of Theorem~\ref{th:1} and Corollary~\ref{co:spacing}}
\label{app:th1}
\begin{IEEEproof}[Proof of Theorem~\ref{th:1}]
	In order to prove the existence and uniqueness of a solution to \eqref{eq:hatCN} for all large $n$, we use the framework of standard interference functions from \cite{YAT95}.
\vspace{0.2cm}

\begin{definition}
	\label{def:standardfunctions}
A function $h=(h_1,\ldots,h_n):\RR_+^n\to \RR_+^n$ is said to be a standard interference function if it fulfills the following conditions:
\begin{enumerate}
	\item {\it Positivity:} if $q_1,\ldots,q_n\geq 0$, then $h_j(q_1,\ldots,q_n)>0$, for all $j$.
	\item {\it Monotonicity:} if $q_1\geq q_1',\ldots,q_n\geq q_n'$, then for all $j$, $h_j(q_1,\ldots,q_n)\geq h_j(q_1',\ldots,q_n')$.
	\item {\it Scalability:} for all $\alpha>1$ and for all $j$, $\alpha h_j(q_1,\ldots,q_n)\geq h_j(\alpha q_1,\ldots,\alpha q_n)$. 
\end{enumerate}
\end{definition}
\vspace{0.2cm}

\vspace{0.2cm}
\begin{theorem}
	\label{th:standardfunctions}
	If an $n$-variate function $h(q_1,\ldots,q_n)$ is a standard interference function and there exists $(q_1,\ldots,q_n)$ such that for all $j$, $q_j\geq h_j(q_1,\ldots,q_n)$, then the system of equations
\begin{equation}
	\label{eq:hj=qj}
q_j = h_j(q_1,\ldots,q_n)
\end{equation}
for $j=1,\ldots,n$, has at least one solution, given by $\lim_{t\to\infty} (q_1^{(t)},\ldots,q_n^{(t)})$, where
\begin{equation*}
	q_j^{(t+1)} = h_j(q_1^{(t)},\ldots,q_n^{(t)})
\end{equation*}
for $t\geq 1$ and any initial values $q_1^{(0)},\ldots,q_n^{(0)}\geq 0$. 
\end{theorem}
\begin{IEEEproof}
	The proof is provided in Appendix \ref{app:standardfunctions}.
\end{IEEEproof}
\begin{remark}
	Note that our definition of a standard interference function differs from that of \cite{YAT95} in which the scalability requirement reads: for all $j$, $\alpha h_j(q_1,\ldots,q_n)>h_j(\alpha q_1,\ldots,\alpha q_n)$. Changing the strict inequality to a loose one alters the consequences for the theorem above, where only existence is ensured. However, for our present purposes with $\phi(s)$ possibly possessing a flat region, requesting a strict inequality would be too demanding.
\end{remark}
\vspace{0.2cm}

Since $\{x_1,\ldots,x_n\}$ spans $\CC^N$ for all large $n$ a.s. (as a consequence of Proposition~\ref{prop:no_eigenvalue} in Appendix~\ref{app:lemmas}), we can define for these $n$ the functions $h_j$, $j=1,\ldots,n$,
\begin{align}
	\label{eq:hj}
	h_j(q_1,\ldots,q_n) \triangleq \frac1Nx_j^*\left(\frac1n\sum_{i=1}^n u(q_i) x_ix_i^*\right)^{-1}x_j.
\end{align}

We first show that $h=(h_1,\ldots,h_n)$ meets the conditions of Theorem~\ref{th:standardfunctions} for all large $n$ a.s. Due to {\bf A1}, from standard arguments using the Markov inequality and the Borel Cantelli lemma, we have that $\min_{i\leq n}\Vert x_i\Vert \neq 0$ for all large $n$ a.s. (this is also a corollary of Lemma~\ref{le:convquadraticform} below). Therefore, we clearly have $h_j>0$ for all $j$, for all large $n$ a.s. Also, since $u$ is non-increasing, taking $q_1,\ldots,q_n$ and $q_1',\ldots,q_n'$ such that $q_i'\geq q_i\geq 0$ for all $i$, $u(q_i')\leq u(q_i)$ and then
\begin{align*}
	\frac1n \sum_{i=1}^n u(q_i)x_ix_i^* \succeq \frac1n\sum_{i=1}^n u(q_i') x_ix_i^* 
\end{align*}
From \cite[Corollary 7.7.4]{HOR85}, this implies
\begin{align*}
	\left(\frac1n \sum_{i=1}^n u(q_i')x_ix_i^*\right)^{-1} \succeq \left(\frac1n\sum_{i=1}^n u(q_i) x_ix_i^*\right)^{-1}
\end{align*}
from which $h_j(q_1',\ldots,q_n')\geq h_j(q_1,\ldots,q_n)$, proving the monotonicity of $h$. 

For $\alpha>1$, $\phi(\alpha q_i)\geq \phi(q_i)$, so that $u(\alpha q_i) \geq \frac{u(q_i)}{\alpha}$. Therefore
\begin{align*}
	\frac1n \sum_{i=1}^n u(\alpha q_i)x_ix_i^* \succeq \frac1{\alpha}\frac1n\sum_{i=1}^n u(q_i) x_ix_i^* 
\end{align*}
From \cite[Corollary 7.7.4]{HOR85} again, we then have
\begin{align*}
	\alpha \left(\frac1n\sum_{i=1}^n u(q_i) x_ix_i^*\right)^{-1} \succeq  \left(\frac1n \sum_{i=1}^n u(\alpha q_i)x_ix_i^*\right)^{-1}
\end{align*}
so that $\alpha h_j(q_1,\ldots,q_n)\geq h_j(\alpha q_1,\ldots,\alpha q_n)$. Therefore $h$ is a standard interference function. In order to prove that \eqref{eq:hj} admits a solution, from Theorem~\ref{th:standardfunctions}, we now need to prove that there exists $(q_1,\ldots,q_n)$ such that for all $j$, $q_j\geq h_j(q_1,\ldots,q_n)$. Note that this may not hold for all fixed $N,n$ as discussed in \cite[pp. 54]{MAR76}. We will prove instead that a solution exists for all large $n$ a.s.

%
%
%

To pursue, we need random matrix results and additional notations. Take $c_-,c_+$ such that $0<c_-<\lim\inf_N c_N$ and $\lim\sup_N c_N<c_+<1$, and denote $X_{(i)}=[x_1,\ldots,x_{i-1},x_{i+1},\ldots,x_n]\in\CC^{N\times (n-1)}$. We start with the following fundamental lemmas, which allow for a control of the joint convergence of the quadratic forms $\frac1Nx_i^*\hat{S}_N^{-1}x_i - 1$.

\vspace{0.2cm}
	\begin{lemma}
		\label{le:lambdamin}
		Assume {\bf A1}--{\bf A3}. There exists $\varepsilon>0$ such that
		\begin{align*}
			\min_{i\leq n} \left\{\lambda_1\left( \frac1nX_{(i)}X_{(i)}^*\right)\right\} &> \varepsilon
		\end{align*}
		for all large $n$ a.s.
	\end{lemma}
	\begin{IEEEproof}
		The proof is provided in Appendix \ref{app:lambdamin}.
	\end{IEEEproof}
\vspace{0.2cm}

\vspace{0.2cm}
\begin{lemma}
	\label{le:convquadraticform}
	Assume {\bf A1}--{\bf A3}. Then, a.s.,
	\begin{align*}
		\max_{i\leq n} \left\{\left| \frac1Nx_i^*\hat{S}_N^{-1}x_i - 1 \right|\right\} \to 0.
	\end{align*}
\end{lemma}
\begin{IEEEproof}
	The proof is provided in Appendix \ref{app:convquadraticform}.
\end{IEEEproof}
\vspace{0.2cm}

Let $q_1=\ldots=q_n\triangleq q>0$. Then, 
\begin{align*}
	h_i(q_1,\ldots,q_n) =\frac1{u(q)} \frac1N x_i^* \hat{S}_N^{-1} x_i = \frac{q}{\phi(q)} \frac1N x_i^* \hat{S}_N^{-1} x_i.
\end{align*}
Take $\varepsilon>0$ such that $(1+\varepsilon)/(\phi_\infty-\varepsilon)<1$. This is always possible since $\phi_\infty>1$. Choose now $q$ such that $\phi(q) = \phi_\infty -\varepsilon$, which also exists since $\phi$ is increasing on $[0,\phi^{-1}(\phi_\infty-))$ with image $[0,\phi_\infty)$. From Lemma~\ref{le:convquadraticform}, for all large $n$ a.s.,
\begin{align*}
	\sup_i \left| \frac1q h_i(q_1,\ldots,q_n)(\phi_\infty-\varepsilon) - 1 \right| < \varepsilon.
\end{align*}
Therefore,
\begin{align*}
	\frac1q h_i(q_1,\ldots,q_n) < \frac{1+\varepsilon}{\phi_\infty-\varepsilon} < 1
\end{align*}
from which $h_i(q,\ldots,q) < q$ for all $i$. From Theorem~\ref{th:standardfunctions}, we therefore prove the existence of a solution to \eqref{eq:hj=qj} with $h_j$ given in \eqref{eq:hj}. Since these quadratic forms define the solutions of the fixed-point equation \eqref{eq:hatCN}, this proves the existence of a solution $\hat{C}_N$ for all large $n$ a.s. Note that Lemma~\ref{le:convquadraticform} is crucial here and that, for $\phi_\infty$ close to one, there is little hope to prove existence for all fixed $N,n$, consistently with the results \cite{MAR76,KEN91}.

We now prove uniqueness. Take a solution $\hat{C}_N$ and denote $d_i = \frac1Nx_i^*\hat{C}_N^{-1}x_i$, which we order as $d_1\leq\ldots\leq d_n$ without loss of generality. Denote also $D=\diag(\{u(d_i)\}_{i=1}^n)$. By definition
\begin{align*}
	d_i = \frac1Nx_i^* \left(\frac1n XDX^*\right)^{-1}x_i.
\end{align*}
From the non increasing property of $u$, we have the inequality
\begin{align*}
XDX^* \succeq u(d_n)XX^*
\end{align*}
which implies after inversion
\begin{align*}
	\frac1{u(d_n)}\left(XX^*\right)^{-1} \succeq \left(XDX^*\right)^{-1}
\end{align*}
and therefore, recalling that $n^{-1}XX^*=\hat{S}_N$, 
\begin{align*}
	d_n \leq \frac1{u(d_n)} \frac1Nx_n^*\hat{S}_N^{-1}x_n
\end{align*}
or equivalently, since $u(d_n)>0$,
\begin{align*}
	\phi(d_n) \leq \frac1Nx_n^*\hat{S}_N^{-1}x_n.
\end{align*}

Similarly,
\begin{align*}
	d_1 \geq \frac1{u(d_1)} \frac1Nx_1^*\hat{S}_N^{-1}x_1
\end{align*}
from which we also have
\begin{align*}
	\phi(d_1) \geq \frac1Nx_1^*\hat{S}_N^{-1}x_1.
\end{align*}

Since $\phi$ is non-decreasing, we also have $\phi(d_1)\leq \phi(d_i) \leq \phi(d_n)$ for $i\leq n$, and we therefore obtain
\begin{align*}
	\frac1Nx_1^*\hat{S}_N^{-1}x_1 \leq \phi(d_i) \leq \frac1Nx_n^*\hat{S}_N^{-1}x_n.
\end{align*}

Take $0<\varepsilon<\min\{1,(\phi_\infty-1)\}$. From Lemma~\ref{le:convquadraticform}, for all large $n$ a.s.,
\begin{align*}
	0<1-\varepsilon < \phi(d_i) < 1+\varepsilon<\phi_\infty.
\end{align*}

Since $\phi$ is continuous and increasing on $(0,\phi^{-1}(\phi_\infty-))$ with image contained in $(0,\phi_\infty)$, $\phi$ is invertible there and we obtain that for all large $n$ a.s.,
\begin{align}
	\label{eq:bound}
	\phi^{-1}\left(1-\varepsilon\right) < d_i < \phi^{-1}\left(1+\varepsilon \right).
\end{align}

We can now prove the almost sure uniqueness of $\hat{C}_N$ for all large $n$. Take $\varepsilon$ in \eqref{eq:bound} to satisfy the previous conditions and to be such that $(\phi^{-1}(1+\varepsilon))^2/\phi^{-1}(1-\varepsilon)<\phi^{-1}(\phi_\infty-)$, which is always possible as the left-hand side expression is continuous in $\varepsilon$ with limit $\phi^{-1}(1)<\phi^{-1}(\phi_\infty-)$ as $\varepsilon\to 0$. 

We now follow the arguments of \cite[Theorem~1]{YAT95}. Assume $(d^{(1)}_1,\ldots,d^{(1)}_n)$ and $(d^{(2)}_1,\ldots,d^{(2)}_n)$ are two distinct solutions of the fixed-point equation $d_j=h_j(d_1,\ldots,d_n)$ for $j=1,\ldots,n$, where $h_j$ is defined by \eqref{eq:hj}. Then (up to a change in the indices $1$ and $2$), there exists $k$ such that, for some $\alpha>1$, $\alpha d^{(1)}_k=d^{(2)}_k$ and $\alpha d^{(1)}_i\geq d^{(2)}_i$ for $i\neq k$. From \eqref{eq:bound}, for sufficiently large $n$ a.s. the ratio $\alpha=d^{(1)}_k/d^{(2)}_k$ is also constrained to satisfy $\alpha<\phi^{-1}(1+\varepsilon)/\phi^{-1}(1-\varepsilon)$. Using this inequality and the upper bound in \eqref{eq:bound}, we have for all $j$ 
\begin{align*}
	0< \alpha d^{(1)}_j < \frac{(\phi^{-1}(1+\varepsilon))^2}{\phi^{-1}(1-\varepsilon)}<\phi^{-1}(\phi_\infty-).
\end{align*}
Since $\phi$ is increasing on $(0,\phi^{-1}(\phi_\infty-))$, we have in particular $\phi(\alpha d_j^{(1)})>\phi(d_j^{(1)})$ from which $\alpha u(\alpha d_j^{(1)})>u(d_j^{(1)})$, for all $j$ and then, with similar arguments as previously, $\alpha h_j(d_1^{(1)},\ldots,d_n^{(1)})>h_j(\alpha d_1^{(1)},\ldots,\alpha d_n^{(1)})$ for all $j$. Using the monotonicity of $h$, we conclude in particular 
\begin{align*}
d^{(2)}_k = h_k(d^{(2)}_1,\ldots,d^{(2)}_n)&\leq h_k(\alpha d^{(1)}_1,\ldots,\alpha d^{(1)}_n) \\
&<\alpha h_k(d^{(1)}_1,\ldots,d^{(1)}_n)=\alpha d^{(1)}_k
\end{align*}
which contradicts $\alpha d^{(1)}_k=d^{(2)}_k$ and proves the uniqueness of $\hat{C}_N$ and Part (I) of Theorem~\ref{th:1}.

\bigskip

We now prove Part (II) of the theorem. In order to proceed, we start again from \eqref{eq:bound}. Since $\varepsilon$ is arbitrary, we conclude that
\begin{align*}
	\max_{i\leq n} \left| d_i - \phi^{-1}(1) \right| \asto 0.
\end{align*}
Applying the continuous mapping theorem, we then have
\begin{align*}
	\max_{i\leq n} \left| u(d_i) -u(\phi^{-1}(1)) \right| \asto 0.
\end{align*}
Noticing that $\phi^{-1}(1) u(\phi^{-1}(1)) = \phi(\phi^{-1}(1))=1$, and therefore that $u(\phi^{-1}(1))=1/\phi^{-1}(1)$, this can be rewritten 
\begin{align}
	\label{eq:uditophi}
	\max_{i\leq n} \left| u(d_i) - \frac1{\phi^{-1}(1)} \right| \asto 0.
\end{align}
Now, we also have the matrix inequalities
\begin{align*}
	&\min_{i\leq n} \left\{u(d_i) - \frac1{\phi^{-1}(1)} \right\}  \frac1nXX^* \\
	&\preceq \frac1n\sum_{i=1}^n \left(u(d_i) - \frac1{\phi^{-1}(1)}\right) x_ix_i^* \\
	&\preceq \max_{i\leq n} \left\{u(d_i) - \frac1{\phi^{-1}(1)} \right\}  \frac1nXX^*.
\end{align*}
From Proposition~\ref{prop:no_eigenvalue} in Appendix~\ref{app:lemmas}, $\Vert \frac1nXX^*\Vert<K$ for some $K>0$ and for all $n$ a.s. From \eqref{eq:uditophi}, we then conclude that
\begin{align*}
	\left\Vert \frac1n\sum_{i=1}^n \left(u(d_i) - \frac1{\phi^{-1}(1)}\right) x_ix_i^* \right\Vert = \left\Vert \hat{C}_N - \frac{\hat{S}_N}{\phi^{-1}(1)} \right\Vert \asto 0
\end{align*}
which completes the proof of Theorem~\ref{th:1}.
\end{IEEEproof}

\begin{IEEEproof}[Proof of Corollary~\ref{co:spacing}]
	The identity follows from \cite[Theorem~4.3.7]{HOR85}, according to which, for $1\leq i\leq N$,
	\begin{align*}
		\lambda_i\left(\hat{S}_N\right) &\leq \lambda_i\left(\phi^{-1}(1)\hat{C}_N\right) + \lambda_N\left(\hat{S}_N-\phi^{-1}(1)\hat{C}_N\right) \\
		\lambda_i\left(\hat{S}_N\right) &\geq \lambda_i\left(\phi^{-1}(1)\hat{C}_N\right) - \lambda_N\left(\hat{S}_N-\phi^{-1}(1)\hat{C}_N\right).
	\end{align*}
	The result follows by noticing that the second term in both right-hand sides tends to zero a.s. according to Theorem~\ref{th:1}.
\end{IEEEproof}

\section{Proof of Lemma~\ref{le:lambdamin}}
\label{app:lambdamin}
If the set of the eigenvalues of $\frac1nX_{(i)}X_{(i)}^*$ is contained within the set of the eigenvalues of $\frac1nXX^*$, then the result is immediate from Proposition~\ref{prop:no_eigenvalue} in Appendix~\ref{app:lemmas}. We can therefore assume the existence of eigenvalues of $\frac1nX_{(i)}X_{(i)}^*$ which are not eigenvalues of $\frac1nXX^*$. By definition, the eigenvalues of $\frac1nX_{(i)}X_{(i)}^*$ solve the equation in $\lambda$
\begin{align*}
	\det\left( \frac1nX_{(i)}X_{(i)}^* - \lambda I_N \right) = 0.
\end{align*}

Take $\lambda$ not to be also an eigenvalue of $\frac1nXX^*$. Then, developing the above expression, we get
\begin{align*}
	&\det\left( \frac1nX_{(i)}X_{(i)}^* - \lambda I_N \right) \\
	&= \det \left(\frac1n XX^* - \frac1nx_ix_i^* -\lambda I_N \right) \\
	&= \det Q(\lambda) \det \left( I_N - Q(\lambda)^{-\oh} \frac1nx_ix_i^* Q(\lambda)^{-\oh}\right) \\
	&= \det Q(\lambda) \left( 1 - \frac1n x_i^* Q(\lambda)^{-1} x_i \right)
\end{align*}
with the notation $Q(\lambda)\triangleq \frac1nXX^*-\lambda I_N$, where we used $\det(I_N+AB)=\det(I_p+BA)$ in the last line, for $A\in\CC^{N\times p}$ and $B\in\CC^{p\times N}$, with $p=1$ here.

Therefore, since $\lambda$ cannot cancel the first determinant,
\begin{align*}
	\frac1n x_i^* Q(\lambda)^{-1} x_i = \frac1n x_i^* \left(\frac1n XX^* -\lambda I_N \right)^{-1} x_i = 1.
\end{align*}

Let us study the function
\begin{align*}
	x\mapsto f_{n,i}(x) \triangleq \frac1n x_i^* \left(\frac1n XX^* - x I_N \right)^{-1} x_i.
\end{align*}

First note, from a basic study of the asymptotes and limits of $f_{n,i}(x)$, that the eigenvalues of $\frac1nX_{(i)}X_{(i)}^*$ are interleaved with those of $\frac1n XX^*$ (a property known as Weyl's interlacing lemma) and in particular that
\begin{equation}
	\label{eq:interleaving}
	\lambda_1\left(\frac1nX_{(i)}X_{(i)}^*\right) \leq \lambda_1\left(\frac1nXX^*\right) \leq \lambda_2\left(\frac1nX_{(i)}X_{(i)}^*\right).
\end{equation}
Since $\lambda_1(\frac1nXX^*)$ is a.s. away from zero for all large $N$ (Proposition~\ref{prop:no_eigenvalue}), only $\lambda_1(\frac1nX_{(i)}X_{(i)}^*)$ may remain in the neighborhood of zero for at least one $i\leq n$, for all large $n$. 

We will show that this is impossible. Precisely, for all large $n$ a.s., we will show that $f_{n,i}(x)<1$ for any $i\leq n$ and for all $x$ in some interval $[0,\xi)$, $\xi>0$, confirming that no eigenvalue of $\frac1n X_{(i)}X_{(i)}^*$ can be found there. For this, we first use the fact that the $f_{n,i}(x)$ can be uniformly well estimated for all $x<0$ through Proposition~\ref{prop:BaiSil95} in Appendix~\ref{app:lemmas} by a quantity strictly less than one. We then show that the growth of the $f_{n,i}(x)$ for $x$ in a neighborhood of zero can be controlled, so to ensure that none of them reaches $1$ for all $x<\xi$. This will conclude the proof.

We start with the study of $f_{n,i}(x)$ on $\RR^-$. From Lemma~\ref{le:MIL},
\begin{align*}
	f_{n,i}(x) = \frac{\frac1n x_i^*\left(\frac1nX_{(i)}X_{(i)}^*-xI_N\right)^{-1}x_i }{1+\frac1n x_i^*\left(\frac1nX_{(i)}X_{(i)}^*-xI_N\right)^{-1}x_i}.
\end{align*}
Define 
\begin{align*}
 \bar{f}_n(x) \triangleq \frac{c_Ne_N(x)}{1+c_Ne_N(x)}
\end{align*}
with $e_N(x)$ the unique positive solution of (see Proposition~\ref{prop:BaiSil95})
\begin{equation}
	e_N(z) = \int \frac{t}{(1+c_Ne_N(z))^{-1}t-z}dF^{C_N}(t).
\end{equation}
Then, with $Q(x)\triangleq\frac1nXX^*-xI_N$, $Q_i(x)\triangleq\frac1nX_{(i)}X_{(i)}^*-xI_N$,
\begin{align}
	\left| f_{n,i}(x) - \bar{f}_n(x) \right| \nonumber 
	&= \left| \frac{\frac1n x_i^*Q_i(x)^{-1}x_i }{1+\frac1n x_i^*Q_i(x)^{-1}x_i} - \frac{c_Ne_N(x)}{1+c_Ne_N(x)} \right| \\
	&\leq \left| \frac1n x_i^*Q_i(x)^{-1}x_i - c_Ne_N(x) \right| \nonumber \\
	&\label{eq:3terms}\leq \left| \frac1n x_i^*Q_i(x)^{-1}x_i - \frac1n \tr C_NQ_i(x)^{-1} \right| \nonumber \\
	&+ \left| \frac1n \tr C_NQ_i(x)^{-1} - \frac1n \tr C_NQ(x)^{-1} \right| \nonumber \\
	&+\left| \frac1n \tr C_NQ(x)^{-1} - c_Ne_N(x) \right|
\end{align}

Using $(a+b+c)^{p}\leq 3^{p}(a^{p}+b^{p}+c^{p})$ for $a,b,c>0$, and $p\geq 1$ (H\"older's inequality), and applying Lemma~\ref{le:trace_lemma}, Lemma~\ref{le:rank1perturbation}, and Proposition~\ref{prop:BaiSil95} to the right-hand side terms of \eqref{eq:3terms}, respectively, with $p=4+\eta/2$, we obtain
\begin{align*}
	\EE\left[ \left|f_{n,i}(x) - \bar{f}_{n}(x) \right|^{4+\frac{\eta}2}\right] \leq \frac{K}{n^{2+\frac{\eta}4}}
\end{align*}
for some constant $K$ independent of $i$, where we implicitly used {\bf A1}. Therefore, using Boole's inequality on the above event for $i\leq n$, and the Markov inequality, for all $\zeta>0$,
\begin{align*}
	&P\left(\max_{i\leq n} \left|f_{n,i}(x) - \bar{f}_{n}(x) \right| > \zeta \right) \\
	&\leq \sum_{i=1}^n P\left(\left|f_{n,i}(x) - \bar{f}_{n}(x) \right| > \zeta \right) < \frac{K}{\zeta^{4+\frac{\eta}2}n^{1+\frac{\eta}4}}.
\end{align*}
The Borel Cantelli lemma therefore ensures, for all $x<0$,
\begin{align}
	\label{eq:fni}
	\max_{i\leq n} \left| f_{n,i}(x) - \bar{f}_{n}(x) \right| \asto 0.
\end{align}

We now extend the study of $f_{n,i}(x)$ to $x$ in a neighborhood of zero. From Proposition~\ref{prop:no_eigenvalue}, $\lambda_1(\frac1nXX^*) > C_-(1-\sqrt{c_+})^2$ for all large $n$ a.s. (recall that $\lim\sup_N c_N<c_+<1$) so that $f_{n,i}(x)$ is well-defined and continuously differentiable on $U=(-\varepsilon,\varepsilon)$ for $0<\varepsilon<C_-(1-\sqrt{c_+})^2$, for all large $n$ a.s. Take $x\in U$. Since the smallest eigenvalue of $\frac1n XX^* - x I_N$ is lower bounded by $C_-(1-\sqrt{c_+})^2-\varepsilon$ for all large $n$, and that
\begin{align*}
	\max_{i\leq n} \left| \frac1n\Vert x_i\Vert^2 - \frac{1}n\tr C_N \right| \asto 0
\end{align*}
(using similar arguments based on the Boole and Markov inequality reasoning as above), we also have that for all large $n$ a.s.
\begin{align*}
	0<f_{n,i}'(x)< \frac{c_+C_+}{( C_-(1-\sqrt{c_+})^2-\varepsilon )^2} \triangleq K'
\end{align*}
where we used $\lim\sup_N\frac{1}n\tr C_N < c_+C_+$.

From this result, along with the continuity of $f_{n,i}$, for $x\in U$ and for all large $n$ a.s.,
\begin{align*}
f_{n,i}(x) < f_{n,i}(-x)+2x K'.
\end{align*}
In particular, for $\xi=\min\{\varepsilon/2,(1-c_+)/(2K')\}$, 
\begin{align}
	\label{eq:fnix}
	f_{n,i}(\xi) < f_{n,i}(-\xi)+(1-c_+).
\end{align}

Since $e_N(0)=1+c_Ne_N(0)$ by definition \eqref{eq:eN}, 
\begin{align*}
	\bar{f}_n(0) = c_N < c_+
\end{align*}
and $\bar{f}_n(x)$ is continuous and increasing on $U$, so that
\begin{align*}
	\bar{f}_n(-\xi) < c_+.
\end{align*}

Recalling \eqref{eq:fni}, we then conclude that, for all large $n$ a.s.
\begin{align*}
	\max_{i\leq n}f_{n,i}(-\xi)<c_+
\end{align*}
which, along with \eqref{eq:fnix}, gives, for all large $n$ a.s.
\begin{align*}
	\max_{i\leq n} f_{n,i}(\xi) < 1.
\end{align*}

Since $f_{n,i}(x)$ is continuous and increasing on $[0,\xi)$, the equation $f_{n,i}(x)=1$ has no solution on this interval for any $i\leq n$, for all large $n$ a.s., which concludes the proof.

\section{Proof of Lemma~\ref{le:convquadraticform}}
\label{app:convquadraticform}

Define $\hat{S}_{N,(i)}=\hat{S}_N-\frac1nx_ix_i^*$ and denote $\hat{S}_{N,(i)}^{-1}$ its inverse when it exists or the identity matrix otherwise. Take $2\leq p\leq 4+\eta/2$ (see {\bf A1}) and $\varepsilon>0$ as in Lemma~\ref{le:lambdamin}. Denoting $\EE_{x_i}$ the expectation with respect to $x_i$ and $\phi_i = 1_{ \{\lambda_1(\hat{S}_{N,(i)})>\varepsilon\}}$,
\begin{align*}
	&\EE_{x_i}\left[\phi_i \left|\frac{\frac1nx_i^*\hat{S}_{N,(i)}^{-1}x_i}{1+\frac1nx_i^*\hat{S}_{N,(i)}^{-1}x_i} - \frac{\frac1n\tr C_N\hat{S}_{N,(i)}^{-1}}{1+\frac1n\tr C_N\hat{S}_{N,(i)}^{-1}}\right|^p\right] \\
	&= \EE_{x_i}\left[\phi_i \left|\frac{\frac1nx_i^*\hat{S}_{N,(i)}^{-1}x_i - \frac1n \tr C_N\hat{S}_{N,(i)}^{-1}}{\left(1+ \frac1nx_i^*\hat{S}_{N,(i)}^{-1}x_i \right)\left(1+ \frac1n\tr C_N\hat{S}_{N,(i)}^{-1} \right)} \right|^p \right] \\
	&\leq \EE_{x_i}\left[\phi_i\left|\frac1nx_i^*\hat{S}_{N,(i)}^{-1}x_i - \frac1n \tr C_N\hat{S}_{N,(i)}^{-1} \right|^p \right].
\end{align*}
Recalling that $x_i=A_Ny_i$ with $y_i$ having independent zero mean and unit variance entries, from Lemma~\ref{le:trace_lemma}, we have
\begin{align*}
	&\EE_{x_i}\left[\phi_i\left|\frac{\frac1nx_i^*\hat{S}_{N,(i)}^{-1}x_i}{1+\frac1nx_i^*\hat{S}_{N,(i)}^{-1}x_i} - \frac{\frac1n\tr C_N\hat{S}_{N,(i)}^{-1}}{1+\frac1n\tr C_N\hat{S}_{N,(i)}^{-1}}\right|^p\right] \\
	&\leq \frac{\phi_iK_p}{n^\frac{p}2} \left[ \left(\frac{\nu_4}n\tr (C_N\hat{S}_{N,(i)}^{-1})^2 \right)^{\frac{p}2}+\frac{\nu_{2p}}{n^{\frac{p}2}}\tr \left( (C_N\hat{S}_{N,(i)}^{-1})^2 \right)^{\frac{p}2}\right] 
\end{align*}
for some constant $K_p$ depending only on $p$, with $\nu_{\ell}$ any value such that $\EE[|y_{ij}|^\ell]\leq\nu_\ell$ (well defined from {\bf A1}). 
Using $\frac1{n^k}\tr A^k\leq (\frac1n\tr A)^k$ for $A\in\CC^{N\times N}$ nonnegative definite and $k\geq 1$, with here $A=(C_N\hat{S}_{N,(i)}^{-1})^2$, $k=p/2$, this gives
\begin{align}
	&\EE_{x_i}\left[\phi_i\left|\frac{\frac1nx_i^*\hat{S}_{N,(i)}^{-1}x_i}{1+\frac1nx_i^*\hat{S}_{N,(i)}^{-1}x_i} - \frac{\frac1n\tr C_N\hat{S}_{N,(i)}^{-1}}{1+\frac1n\tr C_N\hat{S}_{N,(i)}^{-1}}\right|^p\right] \nonumber \\
	&\leq \frac{\phi_i K_p}{n^\frac{p}2} \left(\nu_4^{\frac{p}2} + \nu_{2p} \right) \left(\frac1n\tr (C_N\hat{S}_{N,(i)}^{-1})^2 \right)^{\frac{p}2} \nonumber\\
	&\leq \frac{K_p}{n^{\frac{p}2}} \left(\nu_4^{\frac{p}2}+\nu_{2p}\right) (c_+ C_+^2\varepsilon^{-2})^{\frac{p}2} \triangleq \frac{K'_p}{n^{\frac{p}2}} \label{eq:ineq0} 
\end{align}
where, in \eqref{eq:ineq0}, we used $\tr AB \leq \Vert A\Vert \tr B$ for $A,B\succeq 0$, $\phi_i\leq 1$, $\Vert \hat{S}_{N,(i)}^{-1}\Vert \leq \varepsilon^{-1}$ when $\phi_i=1$, and $\frac1n \tr C_N^2 \leq c_+C_+^2$. 

This being valid irrespective of $X_{(i)}$, we can take the expectation of the above expression over $X_{(i)}$ to obtain
\begin{align*}
	\EE\left[\phi_i\left|\frac{\frac1nx_i^*\hat{S}_{N,(i)}^{-1}x_i}{1+\frac1nx_i^*\hat{S}_{N,(i)}^{-1}x_i} - \frac{\frac1n\tr C_N\hat{S}_{N,(i)}^{-1}}{1+\frac1n\tr C_N\hat{S}_{N,(i)}^{-1}}\right|^p\right] \leq \frac{K'_p}{n^\frac{p}2}.
\end{align*}

Therefore, from Lemma~\ref{le:MIL},
\begin{align*}
	\EE\left[\phi_i\left|\frac1nx_i^*\hat{S}_{N}^{-1}x_i - \frac{\frac1n\tr C_N\hat{S}_{N,(i)}^{-1}}{1+\frac1n\tr C_N\hat{S}_{N,(i)}^{-1}}\right|^p\right] \leq \frac{K'_p}{n^\frac{p}2}.
\end{align*}

Using Boole's inequality on the $n$ events above with $i=1,\ldots,n$, and Markov inequality, for $\zeta>0$, 
\begin{align*} 
	&P\left( \max_{i\leq n} \left\{ \phi_i \left| \frac1nx_i^*\hat{S}_{N}^{-1}x_i - \frac{\frac1n\tr C_N\hat{S}_{N,(i)}^{-1}}{1+\frac1n\tr C_N\hat{S}_{N,(i)}^{-1}} \right| \right\} > \zeta \right) \\
	&\leq \frac{K'_p\zeta^{-p}}{n^{\frac{p}2-1}}.
\end{align*}
Choosing $4<p\leq 4+\eta/2$, the right-hand side is summable. The Borel-Cantelli lemma then ensures that
\begin{align*}
	\max_{i\leq n}  \left\{ \phi_i \left| \frac1nx_i^*\hat{S}_{N}^{-1}x_i - \frac{\frac1n\tr C_N\hat{S}_{N,(i)}^{-1}}{1+\frac1n\tr C_N\hat{S}_{N,(i)}^{-1}} \right| \right\} \asto 0.
\end{align*}
But, from Lemma~\ref{le:lambdamin}, $\min_i \{\phi_i\}=1$ for all large $n$ a.s. Therefore, we conclude
\begin{align}
	\label{eq:maxin}
	\max_{i\leq n}  \left\{ \left| \frac1nx_i^*\hat{S}_{N}^{-1}x_i - \frac{\frac1n\tr C_N\hat{S}_{N,(i)}^{-1}}{1+\frac1n\tr C_N\hat{S}_{N,(i)}^{-1}} \right| \right\} \asto 0.
\end{align}

Since $\hat{S}_{N,(i)}-\varepsilon I_N\succ 0$ for these large $n$, we also have
\begin{align*}
	&\max_{i\leq n}\left| \frac{\frac1n\tr C_N\hat{S}_{N,(i)}^{-1}}{1+\frac1n\tr C_N\hat{S}_{N,(i)}^{-1}} - \frac{\frac1n\tr C_N\hat{S}_{N}^{-1}}{1+\frac1n\tr C_N\hat{S}_{N}^{-1}} \right| \\
	&=\max_{i\leq n} \left| \frac{\frac1n\tr C_N\hat{S}_{N}^{-1} - \frac1n\tr C_N\hat{S}_{N,(i)}^{-1}}{\left(1+\frac1n\tr C_N\hat{S}_{N,(i)}^{-1}\right)\left(1+\frac1n\tr C_N\hat{S}_{N}^{-1}\right)} \right| \leq \frac1n \frac{C_+}{\varepsilon}
\end{align*} 
where, in the last inequality, we used Lemma~\ref{le:rank1perturbation} with $B=C_N$, $A=\hat{S}_{N,(i)}-\varepsilon I_N$ and $x=\varepsilon$, along with the fact that $(1+x)^{-1}\leq 1$ for $x\geq 0$.

From Proposition~\ref{prop:BaiSil95}, since $\lambda_1(\hat{S}_N)\geq \lambda_1(\hat{S}_{N,(i)})>\varepsilon$ for these large $n$ (see \eqref{eq:interleaving}), we also have
\begin{align*}
	\left| \frac1n\tr C_N\hat{S}_{N}^{-1} - \frac{c_N}{1-c_N} \right| \asto 0 
\end{align*}
and thus, from $c_N(1-c_N)^{-1}/(1+c_N(1-c_N)^{-1})=c_N$,
\begin{align*}
	\left| \frac{\frac1n\tr C_N\hat{S}_{N}^{-1}}{1+\frac1n\tr C_N\hat{S}_{N}^{-1}} - c_N \right| \asto 0.
\end{align*}

Putting things together, this finally gives
\begin{align*}
	\max_{i\leq n}  \left\{ \left| \frac1nx_i^*\hat{S}_{N}^{-1}x_i - c_N \right| \right\} \asto 0
\end{align*}
an expression which, since $c_N>c_->0$ for all large $N$, can be divided by $c_N$, concluding the proof.

\section{Proof of Theorem~\ref{th:standardfunctions}}
\label{app:standardfunctions}
	The proof immediately follows from the arguments of \cite{YAT95}. When the scalability assumption is satisfied with strict inequality, the result is exactly \cite[Theorem~2]{YAT95}. When the scalability assumption is reduced to a loose inequality, \cite[Theorem~1]{YAT95} does not hold, and therefore uniqueness cannot be satisfied. Nonetheless, the existence of a solution follows from the proof of \cite[Lemma 1]{YAT95} which does not call for the scalability assumption. Indeed, since there exists $(q_1,\ldots,q_n)$ such that $q_i\geq h(q_1,\ldots,q_n)$ for all $i$, the algorithm 
	\begin{align*}
		q_j^{(t+1)}=h_j(q_1^{(t)},\ldots,q_n^{(t)})
	\end{align*}
	with $q_j^{(0)}=q_j$, satisfies $q_j^{(1)}\leq q_j^{(0)}$ for all $j$. Assuming $q_j^{(t+1)}\leq q_j^{(t)}$ for all $j$, the monotonicity assumption ensures that $q_j^{(t+2)}\leq q_j^{(t+1)}$ which, by recursion, means that $q_j^{(t)}$ is a non-increasing sequence. Now, since $q_j^{(t)}$ is in the image of $h_j$, $q_j^{(t)}>0$ by positivity, and therefore $q_j^{(t)}$ converges to a fixed-point (not necessarily unique). Such a fixed-point therefore exists. Note that \cite[Lemma 2]{YAT95} provides an algorithm for reaching this fixed-point, starting with $q_j^{(0)}=0$ for all $j$.

\section{Proof of Corollary~\ref{co:RG-MUSIC}}
\label{app:RG-MUSIC}
	If $\hat{C}_N$ is replaced by $\hat{S}_N$ in the statement of the result, then Theorem~\ref{co:RG-MUSIC} is exactly \cite[Theorem~2]{MES08c}, which is a direct consequence of \cite[Theorem~3]{MES08b} with some updated remarks on the $\hat\mu_i$ found in the discussion around \cite[Theorem~17.1]{COUbook}. In order to prove Theorem~\ref{co:RG-MUSIC}, we need to justify the substitution of $\hat{S}_N$ by $\hat{C}_N$. First observe that the result is independent of a scaling of $\hat{S}_N$, and therefore we can freely substitute $\hat{S}_N$ by $\phi^{-1}(1)\hat{C}_N$ instead of $\hat{C}_N$. Using the notations of Mestre in \cite{MES08b}, we first need to extend \cite[Proposition~4]{MES08b}. Call $\hat{g}^C_M(z)$ the equivalent of $\hat{g}_M(z)$ designed from the eigenvectors of $\phi^{-1}(1)\hat{C}_N$ instead of those of $\hat{S}_N$ (referred to as $\hat{R}_M$ in \cite{MES08b} with $M$ in place of $N$, and $N$ in place of $n$). Then, on the chosen rectangular contour $\partial \RR^-_y(m)$, both $\hat{g}^C_M(z)$ and $\hat{g}_M(z)$ are a.s. bounded holomorphic functions for all large $N$; this is due to the exact separation \cite[Theorem~3]{COU10b} of the eigenvalues of $\hat{S}_N$ and the fact that Corollary~\ref{co:spacing} ensures the convergence between the eigenvalues of $\phi^{-1}(1)\hat{C}_N$ and of $\hat{S}_N$. 
	
	From \cite[Equation~(29)]{MES08b}, $\hat{g}_M(z)$ consists of the functions $\hat{b}_M(z)$ and $\hat{m}_M(z)$ for which we also call $\hat{b}^C_M(z)$ and $\hat{m}^C_M(z)$ their equivalents for $\phi^{-1}(1)\hat{C}_N$. We need to show that the respective differences of these functions go to zero. From the definition \cite[Equation~(4)]{MES08b} of $\hat{b}_M(z)$, Theorem~\ref{th:1} and the fact that $\left|\frac1N\tr (A^{-1}-B^{-1})\right|\leq \Vert A^{-1}\Vert \Vert B^{-1}\Vert \Vert A-B\Vert $ for invertible $A,B\in\CC^{N\times N}$, we have immediately that
	\begin{align*}
		\sup_{z\in \partial \RR^-_y(m)} \left| \hat{b}_M(z)-\hat{b}^C_M(z)\right| \asto 0.
	\end{align*}
	Similarly, using \cite[Equation~(6)]{MES08b}, and $\left|a^*(A^{-1}-B^{-1})b\right|\leq |a^*b|\Vert A^{-1}\Vert \Vert B^{-1}\Vert \Vert A-B\Vert$ for $a,b\in\CC^N$, we find
	\begin{align*}
		\sup_{z\in \partial \RR^-_y(m)} \left| \hat{m}_M(z)-\hat{m}^C_M(z)\right| \asto 0.
	\end{align*}
By the dominated convergence theorem, this gives
	\begin{align*}
		\oint_{\partial \RR^-_y(m)} \left(\hat{g}^C_M(z)-\hat{g}_M(z)\right) dz \asto 0
	\end{align*}
	which then immediately extends \cite[Proposition~4]{MES08b} to the present scenario. The second step to be proved is that the residue calculus performed in \cite[Equations (32)--(33)]{MES08b} carries over to the present scenario. The poles within the contour $\partial \RR^-_y(m)$ are the $\hat\lambda_k$ and the $\hat\mu_k$ found in the contour. The indices $k$ such that the $\hat\lambda_k$ and $\hat\mu_k$ are within $\partial \RR^-_y(m)$ are the same for $\hat{S}_N$ and $\phi^{-1}(1)\hat{C}_N$ for all large $N$, due to the exact separation property and Corollary~\ref{co:spacing}. This completes the proof.

\section{Useful lemmas and results}
\label{app:lemmas}
\begin{lemma}[A matrix-inversion lemma]
	\label{le:MIL}
	Let $x\in\CC^N$, $A\in\CC^{N\times N}$, and $t\in\RR$. Then, whenever the inverses exist
	\begin{align*}
		x^*\left(A + txx^*\right)^{-1}x = x^*A^{-1}x (1+t x^*A^{-1}x)^{-1}.
	\end{align*}
\end{lemma}

\begin{lemma}[Rank-one perturbation]
	\label{le:rank1perturbation}
	Let $v\in\CC^N$, $A,B\in\CC^{N\times N}$ nonnegative definite, and $x>0$. Then
	\begin{align*}
		\tr B\left(A+vv^*+xI_N\right)^{-1} - \tr B \left(A+xI_N\right)^{-1} \leq x^{-1}\Vert B\Vert.
	\end{align*}
\end{lemma}

\begin{lemma}[Trace lemma] \cite[Lemma~B.26]{SIL06}
	\label{le:trace_lemma}
	Let $A\in\CC^{N\times N}$ be non-random and $y=[y_1,\ldots,y_N]^\trans\in\CC^N$ be a vector of independent entries with $\EE [y_i]=0$, $\EE[|y_i|^2]=1$, and $\EE[|y_i|^\ell]\leq \nu_\ell$ for all $\ell\leq 2p$, with $p\geq 2$. Then,
	\begin{align*}
		\EE\left[\left| y^* Ay - \tr A \right|^p\right]\leq C_p \left( (\nu_4 \tr AA^*)^{\frac{p}2} + \nu_{2p} \tr(AA^*)^{\frac{p}2} \right)
	\end{align*}
	for $C_p$ a constant depending on $p$ only.
\end{lemma}

\begin{proposition}[A random matrix result]
	\label{prop:BaiSil95}
	Let $X=[x_1,\ldots,x_n]\in\CC^{N\times n}$ with $x_i=A_Ny_i$, $A_N\in\CC^{N\times M}$, $M\geq N$, where $y_i=[y_{i1},\ldots,y_{iM}]\in\CC^M$ has independent entries satisfying $\EE[y_{ij}]=0$, $\EE[|y_{ij}|^2]=1$, $\EE[|y_{ij}|^{\ell}]<\nu_\ell$ for all $\ell \leq 2p$ and $C_N\triangleq A_NA_N^*$ is nonnegative definite with $\Vert C_N\Vert<C_+<\infty$. Assume $c_N=N/n$ and $\bar{c}_N=M/N\geq 1$ satisfy $\lim\sup_N c_N<\infty$ and $\lim\sup_N \bar{c}_N<\infty$, as $N,n,M\to\infty$. Then, for $z<0$, and $p>2$,
\begin{align}
	\label{eq:eN_moment}
	\EE\left[\left|\frac1N\tr C_N\left(\frac1nXX^* -zI_N\right)^{-1} - e_N(z)\right|^p\right] \leq \frac{K_p}{N^{\frac{p}2}}
\end{align}
for $K_p$ a constant depending only on $p$, $\nu_{\ell}$ for $\ell\leq 2p$, and $z$, while $e_N(z)$ is the unique positive solution of
\begin{equation}
	\label{eq:eN}
	e_N(z) = \int \frac{t}{(1+c_Ne_N(z))^{-1}t-z}dF^{C_N}(t)
\end{equation}
where $F^{C_N}$ is the eigenvalue distribution of $C_N$. The function $\RR^-\to \RR^+,~z\mapsto e_N(z)$ is increasing.

Moreover, for any $N_0$, as $N,n\to\infty$ with $\lim\sup_N c_N<\infty$, for $z\in\RR\setminus \mathcal S_{N_0}$, where $\mathcal S_{N_0}$ is the union of the supports of the eigenvalue distributions of $\frac1nXX^*$ for all $N\geq N_0$, 
\begin{align}
	\label{eq:conv_eN}
	\frac1N\tr C_N\left(\frac1nXX^* -zI_N\right)^{-1} - e_N(z) \asto 0.
\end{align}
\end{proposition}
\begin{IEEEproof}
	To prove the first part of Proposition~\ref{prop:BaiSil95}, we follow the steps of the proof of \cite{HAC07}. Note first that we can append $A_N$ into an $M\times M$ matrix by adding rows of zeros, without altering the left-hand side of \eqref{eq:eN_moment}. Using the notations of \cite{HAC07}, we consider the simple case where $A_n=0$ and $\sigma_{ij}^n=C^n_{i}$, where $C_i^n$ denotes the $i$-th eigenvalue of $C_N$. Although this updated proof of \cite{HAC07} would impose $C_N$ to be diagonal, it is rather easy to generalize to non-diagonal $C_N$ (see e.g. \cite{COU09,WAG10}). The proof then extends to the non i.i.d. case when using Lemma~\ref{le:trace_lemma} instead of \cite[(B.1)]{HAC07}. The second part follows from the first part immediately for $z<0$. In order to extend the result to $z\in\RR\setminus \mathcal S_{N_0}$, note that both left-hand side terms in \eqref{eq:conv_eN} are uniformly bounded in any compact $\mathcal D$ away from $\mathcal S_{N_0}$ and including part of $\RR^-$, and are holomorphic on $\mathcal D$. From Vitali's convergence theorem \cite{TIT39}, their difference therefore tends to zero on $\mathcal D$, which is what we need. 
\end{IEEEproof}

\begin{proposition}[No eigenvalue outside the support]
	\label{prop:no_eigenvalue}
	Let $X=[x_1,\ldots,x_n]\in\CC^{N\times n}$ with $x_i=A_Ny_i$, $A_N\in\CC^{N\times M}$, where $y_i=[y_{i1},\ldots,y_{iM}]\in\CC^M$ has independent entries satisfying $\EE[y_{ij}]=0$, $\EE[|y_{ij}|^2]=1$ and $\EE[|y_{ij}|^{4+\eta}]<\alpha$ for some $\eta,\alpha>0$, 
$C_N\triangleq A_NA_N^*$ has bounded spectral norm, and $N,n,M\to\infty$ with $\lim\sup_N N/n <1$, and $1\leq \lim\sup_N M/N <\infty$. Let $N_0$ be an integer and $[a,b]\subset\RR\cup \{\pm \infty\}$, $b>a$, a segment outside the closure of the union of the supports $F^{N/n,C_N}$, $N\geq N_0$, with $F^{t,A}$ the limiting support of the eigenvalues of $\frac1n XX^*$ when $C_N$ has the same spectrum as $A$ for all $N$ and $N/n\to t$. Then, for all large $n$ a.s., no eigenvalue of $\frac1nXX^*$ is found in $[a,b]$.
\end{proposition}
\begin{IEEEproof}
	Appending $A_N$ into an $M\times M$ matrix filled with zeros, this unfolds from \cite[Theorem~3]{COU10b} (for which conditions 1)-3) are met), with the supports $F^{N/n,C_N}$ appended with the singleton $\{0\}$. Now, for $A_N\in\CC^{N\times M}$, such that $A_NA_N^*$ is positive definite, zero is not an eigenvalue of $\frac1nXX^*$ for all $N$, a.s., which gives the result. Condition 1) of \cite[Theorem~3]{COU10b} holds here by definition. Condition 3) is obtained by taking $\psi(x)=x^{2+\eta}$. Condition 2) is obtained by taking $z$ a random variable with Pareto distribution $P(z\leq x)=(1-a^{p-1}x^{1-p})1_{x\geq a}$ for $p=5+\eta$ and $a=\alpha^{\frac1{4+\eta}}$; by Markov inequality, 
	\begin{align*}
		\frac1{n_1n_2}\sum_{i\leq n_1,j\leq n_2}P(y_{ij}>x) &\leq \alpha x^{-4-\eta} = P(z>x).
	\end{align*}
	This $z$ has finite $4+\eta$ order moment, which therefore enforces Condition 2).
\end{IEEEproof}

\bibliography{/home/romano/phd-group/papers/rcouillet/tutorial_RMT/book_final/IEEEabrv.bib,/home/romano/phd-group/papers/rcouillet/tutorial_RMT/book_final/IEEEconf.bib,/home/romano/phd-group/papers/rcouillet/tutorial_RMT/book_final/tutorial_RMT.bib,robust_est}

\begin{thebibliography}{10}
\providecommand{\url}[1]{#1}
\csname url@samestyle\endcsname
\providecommand{\newblock}{\relax}
\providecommand{\bibinfo}[2]{#2}
\providecommand{\BIBentrySTDinterwordspacing}{\spaceskip=0pt\relax}
\providecommand{\BIBentryALTinterwordstretchfactor}{4}
\providecommand{\BIBentryALTinterwordspacing}{\spaceskip=\fontdimen2\font plus
\BIBentryALTinterwordstretchfactor\fontdimen3\font minus
  \fontdimen4\font\relax}
\providecommand{\BIBforeignlanguage}[2]{{%
\expandafter\ifx\csname l@#1\endcsname\relax
\typeout{** WARNING: IEEEtran.bst: No hyphenation pattern has been}%
\typeout{** loaded for the language `#1'. Using the pattern for}%
\typeout{** the default language instead.}%
\else
\language=\csname l@#1\endcsname
\fi
#2}}
\providecommand{\BIBdecl}{\relax}
\BIBdecl

\bibitem{SCH86}
R.~Schmidt, ``{Multiple emitter location and signal parameter estimation},''
  \emph{IEEE Transactions on Antennas and Propagation}, vol.~34, no.~3, pp.
  276--280, 1986.

\bibitem{SCH91}
L.~Scharf, \emph{{Statistical Signal Processing: Detection, Estimation and
  Time-Series Analysis}}.\hskip 1em plus 0.5em minus 0.4em\relax Boston, MA,
  USA: Addison-Wesley, 1991.

\bibitem{MES08}
X.~Mestre, ``{On the asymptotic behavior of the sample estimates of eigenvalues
  and eigenvectors of covariance matrices},'' \emph{{IEEE} Transactions on
  Signal Processing}, vol.~56, no.~11, pp. 5353--5368, Nov. 2008.

\bibitem{HUB64}
P.~J. Huber, ``Robust estimation of a location parameter,'' \emph{The Annals of
  Mathematical Statistics}, vol.~35, no.~1, pp. 73--101, 1964.

\bibitem{MAR06}
R.~A. Maronna, D.~R. Martin, and J.~V. Yohai, \emph{Robust Statistics: Theory
  and Methods}, ser. Wiley Series in Probability and Statistics.\hskip 1em plus
  0.5em minus 0.4em\relax John Wiley \& Sons, 2006.

\bibitem{MAR76}
R.~A. Maronna, ``{Robust M-estimators of multivariate location and scatter},''
  \emph{The annals of statistics}, pp. 51--67, 1976.

\bibitem{KEL70}
D.~Kelker, ``{Distribution theory of spherical distributions and a
  location-scale parameter generalization},'' \emph{Sankhy{\=a}: The Indian
  Journal of Statistics, Series A}, vol.~32, no.~4, pp. 419--430, 1970.

\bibitem{OLI12}
E.~Ollila, D.~Tyler, V.~Koivunen, and H.~V. Poor, ``Complex elliptically
  symmetric distributions: survey, new results and applications,'' \emph{{IEEE}
  Transactions on Signal Processing}, vol.~60, no.~11, pp. 5597--5625, 2012.

\bibitem{WAT85}
S.~Watts, ``{Radar Detection Prediction in Sea Clutter Using the Compound
  {K}-Distribution model},'' \emph{IEE Proceeding, Part. F}, vol. 132, no.~7,
  pp. 613--620, December 1985.

\bibitem{SIL95b}
J.~W. Silverstein, ``{Strong convergence of the empirical distribution of
  eigenvalues of large dimensional random matrices},'' \emph{Journal of
  Multivariate Analysis}, vol.~55, no.~2, pp. 331--339, 1995.

\bibitem{MAR67}
V.~A. Mar\u{c}enko and L.~A. Pastur, ``{Distribution of eigenvalues for some
  sets of random matrices},'' \emph{Math USSR-Sbornik}, vol.~1, no.~4, pp.
  457--483, 1967.

\bibitem{SIL98}
Z.~D. Bai and J.~W. Silverstein, ``{No eigenvalues outside the support of the
  limiting spectral distribution of large dimensional sample covariance
  matrices},'' \emph{The Annals of Probability}, vol.~26, no.~1, pp. 316--345,
  1998.

\bibitem{BIA10}
P.~Bianchi, J.~Najim, M.~Maida, and M.~Debbah, ``{Performance of some
  eigen-based hypothesis tests for collaborative sensing},'' \emph{{IEEE}
  Transactions on Information Theory}, vol.~57, no.~4, pp. 2400--2419, 2011.

\bibitem{MES08b}
X.~Mestre, ``{Improved estimation of eigenvalues of covariance matrices and
  their associated subspaces using their sample estimates},'' \emph{{IEEE}
  Transactions on Information Theory}, vol.~54, no.~11, pp. 5113--5129, Nov.
  2008.

\bibitem{COU10b}
R.~Couillet, J.~W. Silverstein, Z.~D. Bai, and M.~Debbah, ``{Eigen-inference
  for energy estimation of multiple sources},'' \emph{{IEEE} Transactions on
  Information Theory}, vol.~57, no.~4, pp. 2420--2439, 2011.

\bibitem{ELK09}
N.~{El Karoui}, ``Concentration of measure and spectra of random matrices:
  applications to correlation matrices, elliptical distributions and beyond,''
  \emph{The Annals of Applied Probability}, vol.~19, no.~6, pp. 2362--2405,
  2009.

\bibitem{FRA08}
U.~J. G.~Frahm, ``Tyler’s m-estimator, random matrix theory, and generalized
  elliptical distributions with applications to finance,'' \emph{Discussion
  papers in statistics and econometrics}, vol.~2, no.~7, 2008.

\bibitem{TYL88}
D.~E. Tyler, ``Some results on the existence, uniqueness, and computation of
  the m-estimates of multivariate location and scatter,'' \emph{SIAM Journal on
  Scientific and Statistical Computing}, vol.~9, p. 354, 1988.

\bibitem{KEN91}
J.~T. Kent and D.~E. Tyler, ``{Redescending M-estimates of multivariate
  location and scatter},'' \emph{The Annals of Statistics}, pp. 2102--2119,
  1991.

\bibitem{MES08c}
X.~Mestre and M.~Lagunas, ``{Modified subspace algorithms for DoA estimation
  with large arrays},'' \emph{{IEEE} Transactions on Signal Processing},
  vol.~56, no.~2, pp. 598--614, Feb. 2008.

\bibitem{YIN88b}
J.~W. Silverstein, Z.~D. Bai, and Y.~Q. Yin, ``{A note on the largest
  eigenvalue of a large dimensional sample covariance matrix},'' \emph{Journal
  of Multivariate Analysis}, vol.~26, no.~2, pp. 166--168, 1988.

\bibitem{COUbook}
R.~Couillet and M.~Debbah, \emph{{Random Matrix Methods for Wireless
  Communications}}, 1st~ed.\hskip 1em plus 0.5em minus 0.4em\relax New York,
  NY, USA: Cambridge University Press, 2011.

\bibitem{YAT95}
R.~D. Yates, ``{A framework for uplink power control in cellular radio
  systems},'' \emph{{IEEE} Journal on Selected Areas in Communications},
  vol.~13, no.~7, pp. 1341--1347, 1995.

\bibitem{HOR85}
R.~A. Horn and C.~R. Johnson, \emph{{Matrix Analysis}}.\hskip 1em plus 0.5em
  minus 0.4em\relax Cambridge University Press, 1985.

\bibitem{SIL06}
Z.~D. Bai and J.~W. Silverstein, \emph{{Spectral analysis of large dimensional
  random matrices}}, 2nd~ed.\hskip 1em plus 0.5em minus 0.4em\relax New York,
  NY, USA: Springer Series in Statistics, 2009.

\bibitem{HAC07}
W.~Hachem, P.~Loubaton, and J.~Najim, ``{Deterministic equivalents for certain
  functionals of large random matrices},'' \emph{Annals of Applied
  Probability}, vol.~17, no.~3, pp. 875--930, 2007.

\bibitem{COU09}
R.~Couillet, M.~Debbah, and J.~W. Silverstein, ``{A deterministic equivalent
  for the analysis of correlated MIMO multiple access channels},'' \emph{{IEEE}
  Transactions on Information Theory}, vol.~57, no.~6, pp. 3493--3514, Jun.
  2011.

\bibitem{WAG10}
\BIBentryALTinterwordspacing
S.~Wagner, R.~Couillet, M.~Debbah, and D.~T.~M. Slock, ``{Large system analysis
  of linear precoding in MISO broadcast channels with limited feedback},''
  \emph{{IEEE} Transactions on Information Theory}, vol.~58, no.~7, pp.
  4509--4537, 2012. [Online]. Available: \url{http://arxiv.org/abs/0906.3682}
\BIBentrySTDinterwordspacing

\bibitem{TIT39}
E.~C. Titchmarsh, \emph{{The Theory of Functions}}.\hskip 1em plus 0.5em minus
  0.4em\relax New York, NY, USA: Oxford University Press, 1939.

\end{thebibliography}

\end{document}